\documentclass[aps,prl,amssymb,groupedaddress,nofootinbib]{revtex4}

\usepackage{epsfig}
\usepackage[english]{babel}

\usepackage{amsfonts}
\usepackage{amssymb}
\usepackage{amsmath}
\usepackage{slashbox}
\usepackage{multirow}

\def\O{\Phi}

\def\={\nonumber &=}

\def\&{{}&}

\def\({\left(}
\def\){\right)}
\def\[{\left[}
\def\]{\right]}
\def\<{\left\langle}
\def\>{\right\rangle}

\def\bk{{\bf k}}

\def\bx{{\bf x}}

\def\curl{\mathcal}

\def\eq{\begin{align}}
\def\qe{\end{align}}
\def\eqa{\begin{eqnarray}}
\def\qea{\end{eqnarray}}

\def\and{\quad \mbox{and} \quad}

\def\bfnl{\kern2pt\overline{\kern-2ptf}_\textrm{NL}}

\def\barQ{\kern2pt\overline{\kern-2pt\curl{Q}}}

\def\barR{\kern2pt\overline{\kern-2pt\curl{R}}}

\def\setsize{\csname @setfontsize\endcsname \setsize}

\begin{document}


\title{Universal Non-Gaussian Initial Conditions for N-body Simulations}

\author{D.M.~Regan$^{1}$}
\author{M.M.~Schmittfull$^{2}$}
\author{E.P.S.~Shellard$^{2}$}
\author{J.R.~Fergusson$^{2}$}

\affiliation
{$^{1}$Department of Physics and Astronomy,
University of Sussex,
Brighton,
BN1 9QH, UK\\
$^{2}$Centre for Theoretical Cosmology,
DAMTP,
University of Cambridge,
 CB3 0WA, UK}

\date{\today}

\begin{abstract}
In this paper we present the implementation of an efficient formalism for the generation of arbitrary non-Gaussian initial conditions for use in N-body simulations. The methodology involves the use of a separable modal approach for decomposing a primordial bispectrum or trispectrum. This approach allows for the far more efficient generation of the non-Gaussian initial conditions already described in the literature, as well as the generation for the first time of non-separable bispectra and the special class of diagonal-free trispectra. The modal approach also allows for the reconstruction of the spectra from given realisations, a fact which is exploited to provide an accurate consistency check of the simulations.
\end{abstract}


\maketitle

\twocolumngrid

\section{Introduction}
Testing for deviations of primordial density fluctuations from Gaussianity represents one of the most active areas of research in cosmology today (see for example \cite{LigSef2010,Crem,0406398,0612571,0701921,Senatore:2009gt,Chen2010, 10036097,DS,Lewis,10015217}). Detection of an appreciable deviation would violate the current slow-roll inflationary paradigm. To date most tests of non-Gaussianity have focussed on constraining the primordial skewness, described by the three-point function or bispectrum, using the cosmic microwave background (CMB). The resultant CMB non-Gaussianity may be simply related to its primordial `seed' via transfer functions. This relationship reflects the fact that the CMB is well described by linear theory. Large scale structure (LSS) as a three dimensional data source, unlike the two dimensional CMB, offers the possibility of a vast improvement in constraining non-Gaussianity. However, a major drawback is the non-linear relation between the primordial density fluctuation and the resulting distribution of structure. For this reason, the investigation of non-Gaussianity using LSS must take a more empirical approach relying on N-body simulations. Owing to the complexity involved in generating non-Gaussian initial conditions, relatively few models have been studied to date. In fact, aside from the local model only the non-Gaussian bispectra of the equilateral and orthogonal shapes have been studied \cite{VW1,VW2}. The implementation in these latter cases involved an extremely computationally expensive algorithm. In this paper we describe an efficient method to  create non-Gaussian initial conditions for arbitrary bispectra and the special class of diagonal-free trispectra. The approach makes use of the separable decomposition of the primordial spectra, which has been exploited to considerable success in the case of the CMB \cite{FS1,FS2,FLS1,RSF1,FLS2,FRS2,FS3}.  In this paper we present a brief overview of the formalism (for a more detailed exposition see ref.~\cite{FRS3}). We detail a non-trivial check of the simulations, verifying the accuracy and consistency of the approach. Finally we summarise our findings.

\section{Algorithm}
In this section we describe briefly the algorithm for the generation of non-Gaussian initial conditions. We assume that the density field is statistically isotropic. Our treatment is universal in that it covers general bispectra and the class of trispectrum models which depend only on the magnitude of its wavenumbers, i.e.\ diagonal-free trispectra. This case covers almost all trispectra discussed to date in the literature, except for the diagonal-dependent local ($\tau_{NL}$) trispectrum.  However, the local $\tau_{NL}$ may be simply generated using the following expansion about two Gaussian fields, $\phi_G$ and $\psi_G$, (where $\langle \phi_G\psi_G\rangle=0$ \cite{SY})\footnote{In the single field case $\psi_G$ is set to zero and $\tau_{NL}=(6f_{NL}/5)^2$. In general $\tau_{NL}$ obeys the inequality $\tau_{NL}> (6f_{NL}/5)^2/2$ \cite{11013636}.}
\begin{align}
\zeta=\phi_G+\psi_G+f_{NL}\left(\phi_{G}^2-\langle \phi_{G}^2 \rangle\right).
\end{align}
The algorithm described here incorporates the generation of an explicit trispectrum in the absence of a bispectrum and vice versa. It should be noted that the bispectrum term also generates an implicit trispectrum. In the case of the local model this `spurious' trispectrum is the $\tau_{NL}$ model described above. Such trispectra may not be desirable in other models
and so should be subtracted out \cite{FRS3}. This issue will be addressed further in a future paper.

\subsection{Bispectrum}
As described in \cite{FRS3}, an arbitrary primordial bispectrum, $B(k_1,k_2,k_3)$, may be simulated by evaluating the quantity
\begin{align}\label{eq:bispecGen}
\Phi^B(\bk)=&\int \frac{d^3\bk' d^3\bk''}{(2\pi)^3}\delta_D(\bk-\bk'-\bk'')\Phi^G(\bk')\Phi^G(\bk'')\nonumber\\
&\times \frac{B(k,k',k'')}{P(k)P(k')+P(k)P(k'')+P(k')P(k'')},
\end{align}
where $\Phi^G$ is a Gaussian random field with the required power spectrum $P(k)$.    This expression, written in convolved form, was 
used to tackle some specific separable bispectrum models in refs.~\cite{VW1,VW2}.  It  is directly 
related to that employed for creating non-Gaussian CMB map simulations \cite{0612571} which was generalised
with modal methods in ref.~\cite{FLS1}.  The modal approach eliminated potential non-Gaussian contributions to the CMB power spectrum. 
Here, however, the power spectra in the denominator must also be symmetrised to mitigate against these spurious effects \cite{VW2}.  
The expression (\ref{eq:bispecGen}) is the natural choice for initial conditions since,  for the local model of inflation, this procedure 
reduces to the usual 
convolution $\Phi^G*\Phi^G$. The primordial perturbation, $\Phi$, given by 
\begin{align}
\Phi=\Phi^G+\frac{F_{NL}}{2}\Phi^B,
\end{align} 
then obeys (in the limit of weak non-Gaussianity) the desired relations
\begin{align}
\langle \Phi(\bk_1)\Phi(\bk_2) \rangle&=(2\pi)^3 \delta_D(\bk_1+\bk_2) P(k_1),\nonumber\\
\langle \Phi(\bk_1)\Phi(\bk_2)\Phi(\bk_3) \rangle&=(2\pi)^3 \delta_D(\sum \bk_i)F_{NL} B(k_1,k_2,k_3).
\end{align}
The direct calculation of initial conditions via this prescription is not efficient in general due to the non-separable form of the integrand on the second line of \eqref{eq:bispecGen}. However, this term may be rewritten in a separable form using the modal techniques described in \cite{FS1,FS2,FLS1}. In particular, we may expand the integrand within (\ref{eq:bispecGen})  in the form
\begin{align}\label{bispDec1}
\frac{B(k,k',k'')}{P(k)P(k')+P(k)P(k'')+P(k')P(k'')}=\nonumber\\
\sum_{rst} \alpha_{rst}^Q q_{r}(k)q_s(k')q_{t}(k''),
\end{align}
where the $q_r$ are one dimensional orthogonal polynomials on the domain of validity of the bispectrum, that is, the tetrahedral region prescribed by the closure condition imposed by the Dirac delta function.  Note that the form of these mode functions $q_r$ is not important - whether polynomial, trigonometric, wavelet, etc - provided they form a complete set (those used in this paper are close to Legendre polynomials and are defined in ref.~\cite{FLS1}).  We may introduce a partial ordering on the indices used in their 3D products and write 
$\sum_{rst} \alpha_{rst}^Q q_{r}(k)q_s(k')q_{t}(k'')=\sum_{n=\{prs\} } \alpha_n^Q q_{\{r}(k)q_s(k')q_{t\}}(k'')$, where $\{\dots\}$ represents the symmetrised quantity\footnote{In what follows we use the compact notation $Q_n(k,k',k'')$ to represent $q_{\{r}(k)q_s(k')q_{t\}}(k'')$.}. The coefficients $\alpha_n^Q$ characterise the specific model under scrutiny. As has been shown in \cite{FLS2}, relatively few modes ($n_{\rm{max}}=\mathcal{O}(30)$) are needed to accurately account for most of the models described in the literature. We also express the Dirac delta function in the form
\begin{align}
\delta_D(\bk)=\frac{1}{(2\pi)^3}\int d^3 \bx e^{i\bk.\bx}.
\end{align}
The bispectrum contribution may now be efficiently evaluated as
\begin{align}\label{eq:phib}
\Phi^B(\bk)=\sum_n \alpha_n^Q q_{\{r}(k) \int d^3\bx e^{i\bk.\bx}M_s(\bx) M_{t\}}(\bx),
\end{align}
where the filtered density perturbations, $M_s(\bx)$, are given by
\begin{align}\label{eq:filter}
M_s(\bx)=\int \frac{d^3 \bk}{(2\pi)^3}\Phi^G(\bk) q_s(k) e^{-i\bk.\bx}.
\end{align}
Thus the evaluation has been reduced to the calculation of a series of fast Fourier transforms. These expressions are to be evaluated in a box corresponding to a maximum wavenumber $k_{\rm{max}}$. Care must be taken to account for unwanted realisations of the discretisation of the Dirac delta function when the wavevectors, $\bk_i$, align. This can be accounted for simply by restricting the range of the wavevectors 
to $|\bk_i| < 2 k_{\rm{max}}/3$ for the calculation of $\Phi^B$. This limitation is more than offset by the vast improvement in numerical speed and accuracy that the modal method offers.

Once the non-Gaussian primordial potential $\Phi^B(\textbf{k})$ is generated it can be translated into the linear density perturbation $\delta_{\textbf{k},z}$ at some initial redshift $z$ using the Poisson equation and transfer function $T(k)$. From $\delta_{\textbf{k},z}$ one can get initial particle positions and velocities for N-body codes using   the Zel'dovich approximation \cite{zeldovich} or second-order Lagrangian perturbation theory \cite{RS98,Sirko}.

\subsection{Trispectrum}
As indicated already, we shall only consider the special class of diagonal-free trispectra in this paper. Such trispectra are given by the following four-point connected correlator
\begin{align}
\langle \Phi(\bk_1)\Phi(\bk_2)\Phi(\bk_1)\Phi(\bk_1)\rangle_c=&(2\pi)^3\delta_D(\bk_1+\bk_2+\bk_3+\bk_4)\nonumber\\
&\times G_{NL}T(k_1,k_2,k_3,k_4).
\end{align}
A primordial perturbation with the correct power spectrum and trispectrum is then given by
\begin{align}\label{gnldef}
\Phi=\Phi^G+\frac{G_{NL}}{6}\Phi^T,
\end{align}
where
\begin{align}
\Phi^T(\bk)&=\int \frac{d^3\bk' d^3\bk''d^3\bk''}{(2\pi)^6}\delta_D(\bk-\bk'-\bk''-\bk''')\nonumber\\
\times& \frac{T(k,k',k'',k''')}{P(k)P(k')P(k'')+{3\,\rm{perms}}}\Phi^G(\bk')\Phi^G(\bk'')\Phi^G(\bk''').
\end{align}
Without the use of separable methods this integral would be intractable in general. However, separable
methods outlined in \cite{RSF1,FRS2} may again be used to greatly simplify the calculation. In particular, we may write
\begin{align}\label{trispDec1}
&\frac{T(k,k',k'',k''')}{P(k)P(k')P(k'')+{3\,\rm{perms}}}=\nonumber\\
&\hspace*{15mm}\sum_m \overline{\alpha}^Q_m \overline{q}_{\{r}(k)\overline{q}_{s}(k')\overline{q}_{t}(k'')\overline{q}_{u\}}(k'''),
\end{align}
where the $\overline{q}_r$ are one dimensional orthogonal polynomials on the domain of validity of the trispectrum, and where $m$ represents the partial ordering $m=\{rstu\}$. In what follows we may refer to the quantity $\overline{Q}_m$ to represent $\overline{q}_{\{r}\overline{q}_{s}\overline{q}_{t}\overline{q}_{u\}}$.The trispectrum contribution now becomes
\begin{align}
\Phi^T(\bk)=\sum_m \overline{\alpha}_m^Q \overline{q}_{\{r}(k) \int d^3\bx e^{i\bk.\bx}\overline{M}_s(\bx)\overline{M}_t(\bx) \overline{M}_{u\}}(\bx),
\end{align}
where the filtered perturbations, $\overline{M}_s(\bx)$, are as in equation \eqref{eq:filter} except for the replacement of $q_s$ by $\overline{q}_s$. Avoidance of unwanted images of the Dirac delta restricts the domain of validity to $k_i < k_{\rm{max}}/2$.

\section{Algorithm validation}
In order to test the accuracy of the algorithm employed it is necessary to establish the convergence of the average of estimators of particular realisations to the expectation value of the estimator. It should be noted that the efficacy of the primordial decomposition has been tested thoroughly in \cite{FLS1,FLS2,FRS2} with accuracy of at least $\mathcal{O}(90-95\%)$ achievable in the case of the bispectrum with $\lesssim\mathcal{O}(30)$ modes and in the case of the trispectrum with $\lesssim\mathcal{O}(50)$ modes. 

\subsection{Bispectrum estimation}
An estimator for the bispectrum is given by \cite{FRS3}
\begin{align}\label{estBisp}
\mathcal{E}=\int \frac{\Pi_{i=1}^3 d^3\bk_i}{(2\pi)^6}&\frac{\delta_D(\bk_1+\bk_2+\bk_3) B(k_1,k_2,k_3)}{P(k_1)P(k_2)P(k_3)}\nonumber\\
&\times [\O_{\bk_1}\O_{\bk_2}\O_{\bk_3}-3\langle \O_{\bk_1}\O_{\bk_2}\rangle \O_{\bk_3}].
\end{align}
The expectation value of this estimator is given by
\begin{align}
\langle\mathcal{E}\rangle=\frac{V}{\pi}\int_{\mathcal{V}_B}dk_1 dk_2 dk_3 \frac{k_1 k_2 k_3 B^2(k_1,k_2,k_3)}{P(k_1)P(k_2)P(k_3)},
\end{align}
where $\mathcal{V}_B$ is the tetrahedral domain allowed by the triangle condition on the wavenumbers $k_i$, and $V$ is a volume factor given by $V=(2\pi)^3\delta_D (\bf{0})$.  Again we expand the theoretical bispectrum in a separable form 
\begin{align}\label{bispDec2}
 \frac{\sqrt{k_1 k_2 k_3} B(k_1,k_2,k_3)}{\sqrt{P(k_1)P(k_2)P(k_3)}}=\sum_n \alpha^{Q'}_n q_{\{r}(k_1)q_s(k_2)q_{t\}}(k_3).
\end{align}
This quantity is different to (\ref{bispDec1}) used for the initial conditions because here it represents an expansion of the 
predicted signal-to-noise for the given bispectrum model.  However, the two sets of expansion coefficients $\alpha^{Q}_n$ and 
$\alpha^{Q'}_n$ can be directly related for any bispectrum using a matrix transformation, so we only need calculate one set 
of coefficients.  The expansion (\ref{bispDec2}) allows us to write the estimator and its expectation value in the form \cite{FRS3}
\begin{align}\label{eq:bispEst}
\mathcal{E}=&\sum_n  \alpha^{Q'}_n \int d^3 \bx [M_r(\bx)M_s(\bx)M_t(\bx)\nonumber\\
&\hspace{20mm}-\langle M_{\{r}(\bx)M_{s}(\bx)\rangle M_{t\}}(\bx)],\\
\langle \mathcal{E}\rangle=&\sum_{n m}  \alpha^{Q'}_n \alpha^{Q'}_m \gamma_{nm},
\end{align}
where
\begin{align}
\gamma_{n m}&=\frac{V}{\pi}\int_{\mathcal{V}_B}dk_1 dk_2 dk_3 Q_n(k_1,k_2,k_3)Q_m(k_1,k_2,k_3),\nonumber\\
M_r(\bx)&=\int \frac{d^3\bk}{(2\pi)^3} \frac{\O_{\bk} q_r(k)}{\sqrt{k P(k)}}e^{i\bk.\bx}.\nonumber
\end{align}
Expressing $\mathcal{E}=\sum_n \alpha^{Q'}_n \beta^{Q'}_n$, with $\beta^{Q'}_n$ defined by equation \eqref{eq:bispEst}, we establish that
\begin{align}
\langle \beta^{Q'}_n\rangle = \sum_{m} \alpha^{Q'}_m \gamma_{nm}.
\end{align}
It is convenient to create an orthonormal set of mode functions $\mathcal{R}_n$ from the product functions $Q_n$ (this may be done using a Gram-Schmidt orthogonalisation using the the inner product $\langle f g\rangle=\int_{\mathcal{V}_B}dk_1 dk_2 dk_3 f(k_1,k_2,k_3) g(k_1,k_2,k_3)$). In terms of these mode functions the consistency relationship may be easily shown to give
\begin{align}\label{eq:bispTest}
\langle \beta^{\mathcal{R}'}_n\rangle = \alpha^{\mathcal{R}'}_n.
\end{align}

\subsection{Trispectrum estimation}
In the case of the diagonal-free trispectrum, the estimator and its expectation value take the form \cite{FRS3}
\begin{align}
\mathcal{E}=&\int \frac{\Pi_{i=1}^4 d^3\bk_i}{(2\pi)^9}\frac{\delta_D(\bk_1+\bk_2+\bk_3+\bk_4) T(k_1,k_2,k_3,k_4)}{P(k_1)P(k_2)P(k_3)P(k_4)}\nonumber\\
&\times [\O_{\bk_1}\O_{\bk_2}\O_{\bk_3}\O_{\bk_4}-6\langle \O_{\bk_1}\O_{\bk_2}\rangle \O_{\bk_3} \O_{\bk_4}\nonumber\\
&\hspace*{30mm}+3\langle \O_{\bk_1}\O_{\bk_2}\rangle\langle \O_{\bk_3}\O_{\bk_4}\rangle],\\
\langle \mathcal{E}\rangle=&\frac{V}{(2\pi)^6}\int_{\mathcal{V}_T}\left(\Pi_{i=1}^4 dk_i k_i\right)\frac{T^2(k_1,k_2,k_3,k_4)}{P(k_1)P(k_2)P(k_3)P(k_4)}\nonumber\\
&\times \left(\sum_i k_i-|\tilde{k}_{34}|-|\tilde{k}_{24}|-|\tilde{k}_{23}|\right),
\end{align}
where $\tilde{k}_{34}=k_1+k_2-k_3-k_4$ and $\mathcal{V}_T$ represent the domain of validity of the wavenumbers $k_i$ as imposed by the Dirac delta function.
Testing for the accuracy of the initial conditions in this case is slightly more involved than the bispectrum test. In particular, we can achieve this by using separable expansions of the following two theoretical quantities,
\begin{align}
\frac{\sqrt{k_1 k_2 k_3 k_4}T(k_1,k_2,k_3,k_4)}{\sqrt{P(k_1)P(k_2)P(k_3)P(k_4)}}&=\sum_n \overline{\alpha}_{1,n}^Q \overline{Q}_n(k_1,k_2,k_3,k_4),\\\label{trispDec2}
\frac{\sqrt{k_1 k_2 k_3 k_4}T(k_1,k_2,k_3,k_4)}{\sqrt{P(k_1)P(k_2)P(k_3)P(k_4)}}&\left(\sum_i k_i-|\tilde{k}_{34}|-|\tilde{k}_{24}|-|\tilde{k}_{23}|\right)\nonumber\\
&=\sum_n \overline{\alpha}_{2,n}^Q \overline{Q}_n(k_1,k_2,k_3,k_4).
\end{align}
The estimator and its expectation value may now be expressed in the form
\begin{align}
\mathcal{E}&=\sum_n \overline{\alpha}_{1,n}^Q \overline{\beta}_n^Q,\\
\langle \mathcal{E}\rangle&=\sum_{nm} \overline{\alpha}_{1,n}^Q \overline{\alpha}_{2,m}^Q \overline{\gamma}_{nm}, 
\end{align}
where 
\begin{align} \label{eq:betaTrisp}
\overline{\beta}_n^Q&=\int d^3 \bx \Big[ \overline{M}_r(\bx)\,\overline{M}_s(\bx)\,\overline{M}_t(\bx)\,\overline{M}_u(\bx)\nonumber\\
&\hspace*{12mm}-6\langle  \,\overline{M}_{\{r}(\bx)\,\overline{M}_s(\bx)\rangle\, \overline{M}_t(\bx)\,\overline{M}_{u\}}\nonumber\\
&\hspace*{12mm}+3\langle  \,\overline{M}_{\{r}(\bx)\,\overline{M}_s(\bx)\rangle \langle\overline{M}_t(\bx)\,\overline{M}_{u\}}\rangle\Big],\\
\overline{\gamma}_{n m}&=\frac{V}{(2\pi)^6}\int_{\mathcal{V}_T}\Pi_{i=1}^4 dk_i \overline{Q}_n(k_1,k_2,k_3,k_4)\nonumber\\
&\hspace*{30mm}\times \overline{Q}_m(k_1,k_2,k_3,k_4),\\
\overline{M}_r(\bx)&=\int \frac{d^3\bk}{(2\pi)^3} \frac{\O_{\bk} \overline{q}_r(k)}{\sqrt{k P(k)}}e^{i\bk.\bx}.
\end{align}
Hence, we establish that
\begin{align}
\langle \overline{\beta}^Q_n\rangle=\sum_m \overline{\alpha}_{2,m}^Q \overline{\gamma}_{nm}\iff \langle \overline{\beta}^{\mathcal{R}}_n\rangle=\overline{\alpha}_{2,m}^{\mathcal{R}},
\end{align}
where the superscript $\mathcal{R}$ refers to coefficients with respect to the orthonormal mode functions $\overline{\mathcal{R}}_n$ created from the product functions $\overline{Q}_n$.

\section{Results}
In order to demonstrate the efficacy of these modal methods we have 
generated non-Gaussian initial conditions for the following bispectrum models (see e.g.\ ref.~\cite{FLS2}): 
 local, equilateral, constant, orthogonal and flattened (non-separable case).
In addition, we have created trispectrum initial conditions  for the local $g_{NL}$ model and 
the equilateral ($c_1$) model, as well as the constant model (see ref.~\cite{FRS2}).
The decomposition of the primordial shapes, as described by equations \eqref{bispDec1} and \eqref{trispDec1} respectively, was calculated first. Using these expansion coefficients the initial conditions were generated. The accuracy of these initial conditions was then tested using the bispectrum and trispectrum estimation techniques described in the previous section. The primordial decompositions against which the initial conditions were compared are described in the case of the bispectrum by eqn~\eqref{bispDec2}, and in the case of the trispectrum by eqn~\eqref{trispDec2}.

\begin{figure}[htp]
\includegraphics[width=0.49\textwidth]{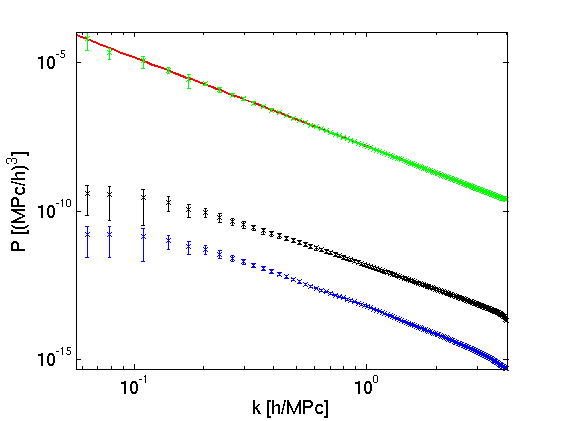}
\caption{Plot of input power $P(k)$ for $\Phi^G$ (red) and measured powers of $f_{NL}\Phi^B/2$ for the local shape with $f_{NL}=100$ (black), $f_{NL}\Phi^B/2$ for the equilateral shape with $f_{NL}=400$ (blue) and $\Phi^G+f_{NL}\Phi^B/2$ for the two cases (green).  $\Phi^B$ was calculated for 100 realisations of $\Phi^G$ using \eqref{eq:phib} on a $256^3$ grid in a $(100\,\rm{Mpc/h})^3$ box.}\label{fig:power}
\end{figure}

\subsection{Bispectrum Results}
The primordial decompositions \eqref{bispDec1} and \eqref{bispDec2} are evaluated using $n_{\rm{max}}=30$ modes in the case of the local, equilateral, constant and orthogonal models, while $n_{\rm{max}}=80$ modes were required in the case of the (non-separable) flattened model in order to achieve a more accurate fit. Since the initial conditions are calculated in the absence of inhomogeneities  we may evaluate the coefficients of the estimator, $\beta_n^Q$, using only the term $\int d^3\bx M_r(\bx) M_s(\bx) M_t(\bx)$ in equation \eqref{eq:bispEst}. 
Evaluation of each initial condition simulation is an extremely efficient operation, with each $M_r$ calculated in parallel.  In the case of a $1024^3$ grid, a full initial condition simulation can be generated in approximately $\mathcal{O}(1)$ hour of computational time using only $6$ cores.   However, the main purpose of this paper is to provide a proof of concept, demonstrating a practical implementation of the methodology presented, so simulations have been carried out using a smaller $256^3$ grid, unless otherwise stated.

Before presenting the bispectrum validation results, we note that the non-Gaussian contribution $f_{NL}^2\langle\Phi^B\Phi^B\rangle/4$ to the power spectrum is small compared to the Gaussian power, as can be seen in  Figure \ref{fig:power} for both local and equilateral models.
This is important because the simulated non-Gaussian contribution in any prescription must not modify the underlying power spectrum. 

\begin{figure}[htp]
\includegraphics[width=0.49\textwidth]{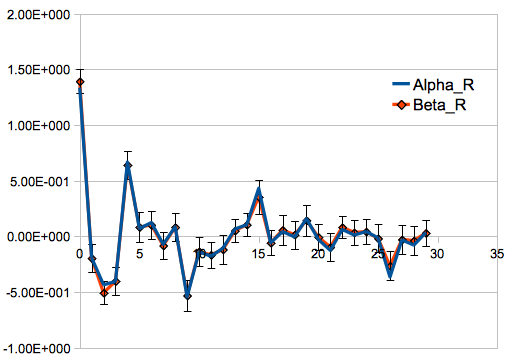}
\caption{Plot of the theoretical local bispectrum modes as described by \eqref{bispDec2} compared to the average of the estimated modes $\beta_n^Q$ from $100$ realisations. The modes are compared in the rotated (orthonormal) $\mathcal{R}$ basis.}\label{fig:bisp}
\end{figure}

In Figure \ref{fig:bisp} we plot a comparison of the theoretical bispectrum modes $\alpha_n^{\mathcal{R}'}$ and the estimated bispectrum modes $\langle\beta_n^{\mathcal{R}}\rangle$ from 100 simulation realizations (1$\sigma$ error bars are shown). The agreement between the theoretical prediction and the averaged simulations is striking. In order to establish the accuracy of the approach we present in Table \ref{tab:num1} the correlations between the primordial shape and the primordial decompositions \eqref{bispDec1} and \eqref{bispDec2}, as well as the correlation between the decomposition \eqref{bispDec2} and the average of the realistions $\beta_n^{Q}$. The amplitude of the bispectrum $F_{NL}$ is also given in the table in each case. It is clear from the table that given a particular decomposition the average of the realisations is almost exact. The only limitation is the number of modes chosen to perform the primordial decompositions. However, in each of the cases considered here, the accuracy of the theoretical decomposition is greater than $90\%$.   It is important to note that the only limitations here have emerged in the theoretical domain, rather than in the simulated initial conditions which faithfully represent the decomposed bispectra they are given by (to better than 99\%).
This theory limitation is easily circumvented by extending out to further coefficients in the modal expansion or by adapting the underlying modes
for the case under investigation. 
\\
\begin{table}[here]
\begin{tabular}{|p{1.1cm}|p{0.7cm}|p{1.0cm}|p{1.75cm}|p{1.75cm}|p{1.4cm}|}
\hline
Model	& $n_{\rm{max}}$ & $F_{NL}$ & Shape vs Decomp Eqn \eqref{bispDec1} & Shape vs Decomp Eqn \eqref{bispDec2}& $\langle\beta_n^Q\rangle$ vs Decomp Eqn \eqref{bispDec2} \\ 	
\hline
Local	&  30 &   100	&  $100\%$	&  $92.6\%$	& $99.5\%$ \\
\hline
Equil	&  30 &   200	&  $99.7\%$	&  $99.7\%$	& $99.8\%$ \\
\hline
Const	&  30 &   200	&  $99.9\%$	&  $100\%$	& $99.3\%$ \\
\hline
Orthog	&  30 &   200	&  $98.7\%$	&  $98.9\%$	& $99.7\%$ \\
\hline
Flat	&  80 &   200	&  $91.8\%$	&  $90.6\%$	& $99.1\%$ \\
\hline
\end{tabular}
\caption{Correlation between the primordial bispectrum shape and the modal decompositions \eqref{bispDec1} and \eqref{bispDec2}, as well as the correlation between the average of the realisations $\langle\beta_n^Q\rangle$ and the decomposition \eqref{bispDec2}.}\label{tab:num1}
\end{table}

\subsection{Trispectrum Results}
\begin{figure}[htp]
\includegraphics[width=0.5\textwidth]{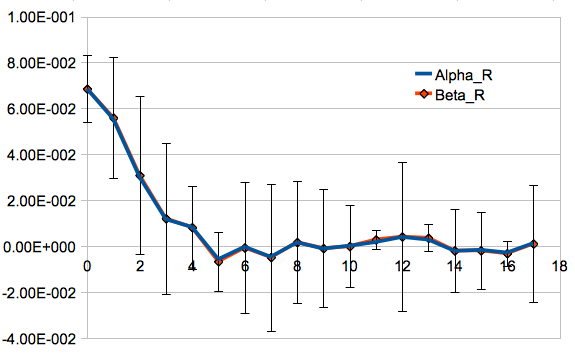}
\caption{Plot of the theoretical modes as described by \eqref{trispDec2} compared to the average of the modes, $\overline{\beta}_n^Q$, over $200$ realisations for the local $g_{NL}$ model. The modes are compared in the rotated (orthonormal) $\overline{\mathcal{R}}$ basis.}\label{fig:trisp}
\end{figure}

The primordial models considered in the case of the trispectrum, i.e. the local ($g_{NL}$), equilateral ($c_1$) and constant models, are quite well-behaved and can be accurately decomposed, as described by equations \eqref{trispDec1} and \eqref{trispDec2}, using just $n_{\rm{max}}=18$ modes. In order to calculate the coefficients $\overline{\beta}_n^Q$ it is necessary to calculate correlators of the form $\langle\overline{M}_r(\bx)\overline{M}_s(\bx)\rangle$, as described by equation \eqref{eq:betaTrisp}. Here, we have carried out $1000$ simulations  in order to measure this correlation accurately. This operation - to test the accuracy of the initial conditions - represents the most numerically intensive operation taking approximately 8 hours on $28$ cores. Again a grid size of $256^3$ is used. The $\overline{\beta}_n^{Q}$ are then calculated and the average over $200$ simulations is calculated and compared to the theoretical prediction \eqref{trispDec2}. In Figure~\ref{fig:trisp} the modes are compared, in the case of the local ($g_{NL}$) model, with $1\sigma$ error bars included for the average of the $\overline{\beta}_n^{\mathcal{R}}$. Clearly, the two sets of modes are highly correlated, validating the trispectrum methodology. In Table~\ref{tab:trisp} we present the correlations between the primordial shape and the primordial decompositions \eqref{trispDec1} and \eqref{trispDec2}, as well as the correlation between the decomposition \eqref{trispDec2} and the average of the realisations $\beta_n^{Q}$, with the amplitude of the trispectrum given by $G_{NL}$ (see equation \eqref{gnldef}). The almost $100\%$ correlation in each case verifies the accuracy and validity of the approach. It should be noted that while the choice of $G_{NL}$ is arbitrary, a lower amplitude will, of course, result in a lower signal to noise for the estimator.

\begin{table}[here]
\begin{tabular}{|p{1.1cm}|p{0.7cm}|p{1.0cm}|p{1.75cm}|p{1.75cm}|p{1.4cm}|}
\hline
Model	& $n_{\rm{max}}$ & $G_{NL}$ & Shape vs Decomp Eqn \eqref{trispDec1} & Shape vs Decomp Eqn \eqref{trispDec2}& $\langle\overline{\beta}_n^Q\rangle$ vs Decomp Eqn \eqref{trispDec2} \\ 	
\hline
Local	&  18 &  $5\times 10^6$ 	&  $99.6\%$	&  $100\%$	& $100\%$ \\
\hline
Equil	&  18 &   $5\times 10^6$ 	&  $99.4\%$	&  $99.8\%$	& $100\%$ \\
\hline
Const	&  18 &   $5\times 10^6$ 	&  $99.9\%$	&  $99.9\%$	& $100\%$ \\
\hline
\end{tabular}
\caption{Correlation between the primordial shape and the decompositions \eqref{bispDec1} and \eqref{bispDec2}, as well as the correlation between the average of the realisations $\langle\beta_n^Q\rangle$ and the decomposition \eqref{bispDec2}.}\label{tab:trisp}
\end{table}

\section{Summary}
We have described the generation of non-Gaussian initial conditions for use in N-body simulations. Exploiting the use of a separable modal approach to analyse primordial non-Gaussianity, the algorithm is reduced to a series of fast Fourier transforms and a three dimensional integral. We have described an application of the approach to arbitrary bispectra, presenting for brevity the implementation of the local, equilateral, constant and orthogonal models, as well as the (non-separable) flattened model. We have also presented the implementation for the class of diagonal-free trispectra, including equilateral ($c_1$), local ($g_{NL}$) and constant models. If primordial bispectra and trispectra
are to be simulated together, then it may be necessary for a spurious primordial bispectrum contribution to the trispectrum to be subtracted out (except in the simplest local $\tau_{NL}$ model). However, the subtraction of such terms involves extending the work presented here to `diagonal-dependent' trispectra.  We defer a detailed quantitative analysis of this trispectrum issue to a future publication. Nonetheless, the present work represents a significant step forward in opening up the efficient investigation of 
primordial non-Gaussianity using large scale structure.
\par
For the non-Gaussian initial conditions described in this paper, a non-trivial consistency check has been carried out to verify their efficacy. It has been established that - once unwanted images of the Dirac delta function are accounted for - the algorithm employed here is unbiased and accurate.

\section*{Acknowledgements}
We thank Hiro Funakoshi for identifying the issue of images of the Dirac delta function.  We are grateful to Andrey Kaliazin for his invaluable computational help.
Simulations were performed on the COSMOS supercomputer (an SGI UV Altix) which is funded by STFC and DBIS. DMR was supported by the STFC grant ST/I000976/1. MMS is supported by an STFC and DAMTP, University of Cambridge. EPS and JRF were supported by the STFC grant ST/F002998/1 and the Centre for Theoretical Cosmology.

\bibliography{LSS}

\end{document}